\documentclass[amsmath,amssymb,twocolumn,11pt,prd,reprint,showpacs,aps]{revtex4-1}

\usepackage{amsmath}
\usepackage{amssymb}
\usepackage{amsthm}
\usepackage{psfrag}
\usepackage{epsfig}
\usepackage{calc}
\usepackage{graphicx}
\usepackage{pgf}
\usepackage{tikz}
\usetikzlibrary{decorations.markings}

\newcommand{\beq}{\begin{equation}}
\newcommand{\eeq}[1]{\label{#1}\end{equation}}
\newcommand{\bea}{\begin{eqnarray}}
\newcommand{\eea}[1]{\label{#1}\end{eqnarray}}
\newcommand{\changed}[1]{#1}
\def\lhl{l\!\cdot\!h\!\cdot\!l}
\def\lh{l\!\cdot\!h}

\begin{document}

\title{\Large\bf Causality Constraints on Massive Gravity}
\author{\textbf{Xi\'an O.~Camanho$^1$, Gustavo Lucena G\'omez$^1$ and Rakibur Rahman$^{1,2}$}\vspace{5pt}}
\affiliation{$^1$Max--Planck--Institut f\"ur Gravitationsphysik (Albert--Einstein--Institut), Am M\"uhlenberg 1, D-14476 Potsdam-Golm, Germany
\\$^2$\text{Department of Physics, University of Dhaka, Dhaka 1000, Bangladesh}\\
\vspace{-5pt}
\\\emph{\texttt{\text{xcamanho@gmail.com,~glucenag@aei.mpg.de,~rakibur.rahman@aei.mpg.de}}}}

\begin{abstract}

\changed{The de Rham--Gabadadze--Tolley massive gravity admits pp-wave backgrounds on which linear fluctuations are shown to undergo time advances for all values of the parameters. The perturbations may propagate in closed time-like curves unless the parameter space is constrained to a line. These classical phenomena take place well within the theory's validity regime.}

\end{abstract}

%\pacs{~}

\maketitle

%%%%%%%%%%%%%%%%%%%
\section{INTRODUCTION}
%%%%%%%%%%%%%%%%%%%

A non-zero graviton mass is an interesting theoretical possibility that modifies General Relativity in the infrared. It is not so easy, though, to construct consistent theories of massive gravity. Such attempts were initiated long ago by Fierz and Pauli~\cite{Fierz:1939ix}, who wrote down
a ghost-free linearized Lagrangian for a massive graviton in flat space. However, it was not until recently that a consistent non-linear theory could be
constructed~\cite{deRham:2010ik,deRham:2010kj}, thanks to de Rham, Gabadadze and Tolley (dRGT). The dRGT massive gravity is remarkable in that it overcame the Boulware--Deser
ghost problem~\cite{Boulware:1973my}, formerly believed to plague any non-linear theory of massive gravity with instabilities.

In this paper, we consider 4D massive gravity theories that admit Minkowski space as a solution. They constitute a family of Lagrangians that include
the graviton mass $m$ as well as two dimensionless parameters $\alpha_3$ and $\alpha_4$:
\beq\mathcal L=\tfrac{1}{2}M_\textrm{P}^2\sqrt{-g}\left[R+m^2\left(\mathcal U_2+\alpha_3\mathcal U_3+\alpha_4\mathcal U_4\right)\right],\eeq{Lagrangian}
where the three possible potential terms are
\beq
\begin{aligned}
\mathcal U_2&=\left[\mathcal K\right]^2-\left[\mathcal K^2\right]\!,\\
\mathcal U_3&=\left[\mathcal K\right]^3-3\left[\mathcal K\right]\left[\mathcal K^2\right]+2\left[\mathcal K^3\right]\!,\\
\mathcal U_4&=\left[\mathcal K\right]^4\!\!-\!6\!\left[\mathcal K\right]^2\!\left[\mathcal K^2\right]\!+\!8\!\left[\mathcal K\right]\!\left[\mathcal K^3\right]
\!+\!3\!\left[\mathcal K^2\right]^2\!\!-\!6\!\left[\mathcal K^4\right]\!,\end{aligned}\eeq{potentials}
with the notation $\left[X\right]\equiv X^\mu{}_\mu$ for the tensor
\beq \mathcal{K}^\mu{}_\nu=\delta^\mu_\nu-\sqrt{g^{\mu\rho}f_{\rho\nu}} \eeq{K-defined}
and its various powers, where $f_{\mu\nu}$ is the reference metric which we will assume to be flat: $f_{\mu\nu}=\eta_{\mu\nu}$.

Eqs.~(\ref{Lagrangian})--(\ref{K-defined}) present the theory in the so-called unitary gauge, in which a Hamiltonian analysis has confirmed the non-existence
of the ghost at the full non-linear level for generic values of the parameters~\cite{Hassan:2011hr}. One may wonder if the absence of unphysical
modes in a theory guarantees its classical consistency. After all, there are various known instances where this is not true~\cite{Velo:1972rt,Henneaux:2013vca,Camanho:2014apa}.
In the context of massive gravity, this issue was raised and critically addressed already in~\cite{Deser:2012qx,Hassan}.

The purpose of this paper is to argue that the dRGT theory may exhibit causality violation well below the strong-coupling scale $\Lambda=\sqrt[3]{m^2M_\textrm{P}}$.
More precisely, the theory admits pp-wave backgrounds that let the longitudinal modes of massive-gravity fluctuations undergo measurable time advances \changed{everywhere 
in parameter space. Should this alone not be considered as a pathology, it is further argued that perturbations may follow closed time-like curves except on the line
\beq \alpha_3=-\tfrac{1}{2}\,.\eeq{mainresult}}

%%%%%%%%%%%%%%%%%%%%%%%
\section{pp-WAVE SOLUTIONS}
%%%%%%%%%%%%%%%%%%%%%%%

Let us introduce the light-cone coordinate system $\left(u,v,\vec{x}\right)$, where
$u=t-x_3$, $v=t+x_3$, and $\vec{x}=\left(x_1,x_2\right)$. In these coordinates, a generic pp-wave spacetime has the following metric:
\beq \mathrm{d}s^2=-\mathrm{d}u\mathrm{d}v+F(u,\vec{x})\mathrm{d}u^2+\mathrm{d}\vec{x}^2\,.\eeq{metric0}
This geometry enjoys the null Killing vector $\partial_v$. One can introduce a covariantly constant null vector $l_\mu=\delta_{\mu u}$
to write this metric in the Kerr--Schild form,
\beq \bar g_{\mu\nu}=\eta_{\mu\nu}+Fl_\mu l_\nu\,.\eeq{metric1}

To see if massive gravity admits  pp-waves solutions, let us first write down the equations of motion resulting from the
Lagrangian~(\ref{Lagrangian}). They are
\beq G_{\mu\nu}+m^2\mathcal X_{\mu\nu}=0\,,\eeq{eom0}
where $G_{\mu\nu}$ is the Einstein tensor and $\mathcal X_{\mu\nu}$ is given by
\beq\begin{aligned}
\!\!\!\!\mathcal X_{\mu\nu}&=\mathcal{K}_{\mu\nu}-\left[\mathcal{K}\right]g_{\mu\nu}\\
&-\alpha\left(\mathcal{K}^2_{\mu\nu}-\left[\mathcal{K}\right]\mathcal{K}_{\mu\nu}+\tfrac{1}{2}g_{\mu\nu}
\big(\left[\mathcal{K}\right]^2-\left[\mathcal{K}^2\right]\big)\right)\\
&-\beta\left(\mathcal{K}^3_{\mu\nu}-\left[\mathcal{K}\right]\mathcal{K}^2_{\mu\nu}+\tfrac{1}{2}
\mathcal{K}_{\mu\nu}\big(\left[\mathcal{K}\right]^2-\left[\mathcal{K}^2\right]\big)\right)\\
&+\tfrac{1}{6}\beta g_{\mu\nu}\big(\left[\mathcal{K}\right]^3-3\left[\mathcal{K}^2\right]\left[\mathcal{K}\right]^2
+2\left[\mathcal{K}^3\right]\big)\,,\end{aligned}\eeq{X-defined}
while the parameters $\alpha$ and $\beta$ are given in terms of the original ones as
$\alpha\equiv3\alpha_3+1,~\beta\equiv-3\left(\alpha_3+4\alpha_4\right)$.

The metric~(\ref{metric1}) yields the following Einstein tensor: $G_{\mu\nu}=-\tfrac{1}{2}l_{\mu}l_{\nu}\partial^2 F$, with $\partial^2\equiv\partial_\mu\partial^\mu$.
To compute $\mathcal X_{\mu\nu}$, note that $\mathcal{K}^\mu{}_\nu=\delta^\mu_\nu-\sqrt{\bar g^{\mu\rho}\left(\bar g_{\rho\nu}-Fl_{\rho}l_\nu\right)}
=\tfrac{1}{2}Fl^\mu l_\nu$.
Since $l_\mu$ is null, $\left[\mathcal K\right]=0$ and $\mathcal K^2_{\mu\nu}=0$. It is then clear from
Eq.~(\ref{X-defined}) that $\mathcal X_{\mu\nu}=\mathcal{K}_{\mu\nu}=\tfrac{1}{2}Fl_\mu l_\nu$.
Therefore, the metric~(\ref{metric1}) will be a solution of the massive gravity equation~(\ref{eom0}),
provided that the function $F$ satisfies the massive Klein--Gordon equation. The latter actually reduces to
the 2D screened Poisson equation since $F=F(u,\vec x)$ is independent of $v$:
\beq \left(\partial^2-m^2\right)F=\left(\partial_i\partial_i-m^2\right)F=0\,.\eeq{Poisson}

Assuming rotational symmetry on the transverse plane, this equation has the following solution at $\vec x \neq 0$:
\beq F=A(u)K_0(m|\vec{x}|)\,,\eeq{sol1}
where $K_0$ is the zeroth-order modified Bessel function of the second kind, while $A(u)$ is arbitrary in $u$.

Let us choose the profile of a ``sandwich wave'' displayed in Fig.~1:
\begin{align} A(u)=\begin{cases}a\exp\left[-\frac{\lambda^2 u^2}{(u^2-\lambda^2)^2}\right]&\text{if}~u\in[-\lambda,\lambda]\,,\label{shock-F}\\
0&\text{otherwise}\,,\end{cases}\end{align}
where $a$ is a numerical constant and $\lambda$ is a length scale~\cite{Bondi:1958aj}. Eq.~(\ref{shock-F}) defines a smooth function $A(u)\in \mathrm{C}^{\infty}(\mathbb{R})$,
with a compact support $[-\lambda,\lambda]$. The sandwich wave moves at the speed of light in the $v$-direction. Its amplitude and width are defined by $a$ and $\lambda$ respectively.

One might wonder about the singularity of the metric at $|\vec x|=0$. In fact, such a geometry may be viewed as arising from
the stress-energy tensor \beq T_{\mu\nu}=\pi M_\textrm{P}^2A(u)\delta^2(\vec x)\,l_\mu l_\nu\,,\eeq{stress-energy}
which saturates the null-energy condition. Then, the energy $E$ of the source is quantified by $M_\textrm{P}^2\lambda$.
%\vspace{-5pt}
\begin{center}
\begin{figure}[ht]
\hspace*{-10pt}
\begin{minipage}{.2\textwidth}
\includegraphics[width=1\linewidth,height=.65\linewidth]{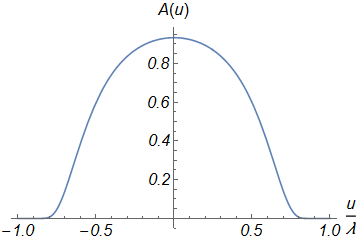}
\end{minipage}
\begin{minipage}{.05\textwidth}
\hphantom{hhhh}
\end{minipage}
\begin{minipage}{.2\textwidth}
\includegraphics[width=1\linewidth,height=.65\linewidth]{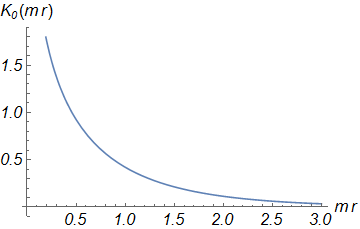}
\end{minipage}
\caption{Profiles of the sandwich wave: $u$-direction profile (left) and radial profile in the transverse plane (right).}
\end{figure}
\end{center}
%\vspace{-20pt}
For future convenience, we choose the amplitude $a$ such that $\int_{-\lambda}^{+\lambda}\mathrm{d}u\,A(u)=\lambda$.
This amounts to the choice $a\approx0.93=\mathcal{O}(1)$. We also choose the width $\lambda$ to be larger than the resolution length of the effective field theory:
$\lambda\gtrsim1/\Lambda$. The latter choice is possible for a very large energy of the source: $E\gg\Lambda$. This situation is completely acceptable and
does not at all invalidate the effective field theory description \cite{footnote1}.

%%%%%%%%%%%%%%%%%%%%%%%%%
\section{LINEAR FLUCTUATIONS}
%%%%%%%%%%%%%%%%%%%%%%%%%

On the pp-wave geometry described in the previous section, let us consider linear massive-gravity fluctuations, $h_{\mu\nu}=g_{\mu\nu}-\bar{g}_{\mu\nu}$.
Schematically, their equations of motion read~\cite{Bernard:2014bfa}
\beq \delta G_{\mu\nu}+m^2\delta \mathcal X_{\mu\nu}=0\,,\eeq{eom-fluc}
with the quantity $\delta G_{\mu\nu}$ given by
\begin{align}\!\!\!\!\delta G_{\mu\nu}&=-\tfrac{1}{2}\left(\nabla^2h_{\mu\nu}-2\nabla^\rho\nabla_{(\mu}h_{\nu)\rho}
+\nabla_\mu\nabla_\nu h\right)\notag\\&+\tfrac{1}{2}\bar g_{\mu\nu}\!
\left(\nabla^2h-\nabla\!\cdot\!\nabla\!\cdot\!h+\bar R^{\rho\sigma}h_{\rho\sigma}\right)-\tfrac{1}{2}
\bar Rh_{\mu\nu}\,,\label{del-G}\end{align}
where $\nabla_{\mu}$ is the covariant derivative built from the background metric $\bar{g}_{\mu\nu}$, dot denotes a contraction of indices and $h\equiv\bar g^{\mu\nu}h_{\mu\nu}$.
To find an expression for $\delta \mathcal{X}_{\mu\nu}$, we first need the variation of the tensor $\mathcal{K}^\mu{}_\nu$ defined in Eq.~(\ref{K-defined}).
An explicit computation gives
\beq \hspace*{-5pt}\delta\mathcal{K}^\mu{}_\nu=\tfrac{1}{2}h^\mu_\nu+\tfrac{1}{8}F\left(l^\mu\lh_\nu-3l_\nu\lh^\mu\right)
-\tfrac{1}{16}F^2l^\mu l_\nu\lhl.\eeq{var-K1}
Then, varying Eqs.~(\ref{X-defined}) one finds from a straightforward calculation that
\begin{align}\!\!\!\!\!\delta\mathcal{X}_{\mu\nu}&=\tfrac{1}{2}\left(h_{\mu\nu}-\bar{g}_{\mu\nu}h\right)-\tfrac{2\alpha-1}{4}
Fl_{(\mu}\lh_{\nu)}+\tfrac{\alpha}{4} Fl_\mu l_\nu h\notag\\&+\left(\tfrac{\alpha+1}{4}\bar{g}_{\mu\nu}
-\tfrac{1}{16}F l_\mu l_\nu\right)F\lhl\,.\label{var-X}\end{align}
Note that the parameter $\beta$ has dropped out! In other words, the linearized fluctuations on the pp-wave
background are insensitive to $\beta$. The subsequent analysis therefore holds for any value of this parameter.

One can now proceed to derive the scalar and vector constraints. We would not bore the reader with the tedious details,
and just present the final results. The trace constraint reads
\beq h=\left(\alpha+\tfrac{1}{2}\right)\left[F\lhl+\tfrac{2}{3m^2}F_{,\:\!\mu}\left(\partial^\mu\lhl-l\!\cdot\!\partial\,
\lh^\mu\right)\right]\,,\eeq{trace1}
whereas the divergence constraint is given by
\begin{align}\!\!\mathcal{C}_\mu&=-\tfrac{\alpha}{2}F\partial_\mu\lhl+\tfrac{2\alpha-1}{4}Fl\!\cdot\!\partial\,\lh_\mu
+\tfrac{2\alpha+3}{4}F_{,\:\!\rho}\,l_\mu\lh^\rho\notag\\&+\tfrac{1}{4}Fl_\mu l\!\cdot\!\partial\,h
-\left(\tfrac{\alpha+1}{2}F_{,\:\!\mu}+\tfrac{2\alpha+1}{16}F^2l_\mu l\!\cdot\!\partial\right)\lhl\,,\label{div-cons}\end{align}
where $\mathcal{C}_\mu\equiv\nabla\!\cdot\!h_\mu-\nabla_\mu h=\partial\!\cdot\!h_\mu-\partial_\mu h-\tfrac{1}{2}F_{,\:\!\mu}\lhl$, and
$F_{,\:\!\mu}$ is a shorthand notation for $\partial_\mu F$.

The derivation relies on the assumption that the fluctuations do not propagate through $\vec x=0$, so that the background equation~(\ref{Poisson})
can be used. Note that it involves not just the divergences and trace of Eq.~(\ref{eom-fluc}), but also contractions thereof with the
null vector $l^\mu$. Of particular interest is the quantity $\nabla^\mu\delta{G}_{\mu\nu}$,
which actually reduces to terms containing only single derivatives of the fluctuations, thanks to the identity~(5.3) of Ref.~\cite{Bernard:2014bfa}.
Also, one needs the background Riemann tensor, which reads $\bar R^\rho{}_{\sigma\mu\nu}=l_\sigma l_{[\mu}\partial_{\nu]}\partial^\rho{F}-l^\rho
l_{[\mu}\partial_{\nu]}\partial_\sigma{F}$.

The 5 constraints~(\ref{trace1})--(\ref{div-cons}) render non-dynamical 5 components of the symmetric tensor $h_{\mu\nu}$,  leaving one with 5
dynamical degrees of freedom, as expected. To be more explicit, we rewrite the scalar constraint as
\beq h=4F\hat{\alpha}h_{vv}+\tfrac{1}{\hat{m}^2}\,F_{,\:\!i}\left(\partial_ih_{vv}-\partial_vh_{vi}\right)\,,\eeq{trace2}
where we have defined $\hat\alpha\equiv\alpha+\tfrac{1}{2}$, and $\tfrac{1}{\hat{m}^2}\equiv\tfrac{8}{3m^2}\hat\alpha$.
Because $h=\left(h_{11}+h_{22}\right)-4\left(h_{uv}+Fh_{vv}\right)$, Eq.~(\ref{trace2}) determines completely the linear
combination $\left(h_{11}+h_{22}\right)$ in terms of other components.
On the other hand, the vector constraint~(\ref{div-cons}) lets one set the 4 components $h_{u\mu}$ to be non-dynamical, since their $v$-derivatives are
completely determined. Therefore, the dynamical degrees of freedom are the two transverse modes: $\left(h_{11}-h_{22}\right)$ and $h_{12}$,
plus the three longitudinal ones: $h_{vi}$ and $h_{vv}$.

To study the true dynamics, let us use commutators of covariant derivatives to rewrite Eq.~(\ref{eom-fluc}) as
\beq \left(\nabla^2-m^2\right)h_{\mu\nu}=\Delta\mathcal{R}_{\mu\nu}\,,\eeq{del-G10}
where the right hand side is written solely in terms of the constraints and curvatures, and is given by
\begin{align}
&\Delta\mathcal{R}_{\mu\nu}=2\nabla_{(\mu}\mathcal{C}_{\nu)}\!+\nabla_\mu\nabla_\nu h-2F_{,\:\!\rho(\mu}l_{\nu)}\lh^\rho\!+F_{,\:\!\mu\nu}\lhl\notag\\
&-\bar g_{\mu\nu}\!\nabla\!\cdot\mathcal C-m^2\bar g_{\mu\nu}\!\left[h-\tfrac{1}{4}(2\hat\alpha-1)F\lhl\right]+\cdots\,,\label{del-G1}
\end{align}
with ellipses standing for terms that do not contribute to the physical modes.
%To proceed further, we need some explicit expressions for the covariant derivatives:
%
%\bea &\nabla^2h_{\mu\nu}=\partial^2h_{\mu\nu}-2F_{,\:\!(\mu}l\!\cdot\!\partial\,\lh_{\nu)}+\cdots\,,&\nonumber\\
%&\nabla_{(\mu}\mathcal{C}_{\nu)}=\partial_{(\mu}\mathcal{C}_{\nu)}+\cdots\,,\qquad
%\nabla_{\mu}\nabla_{\nu}h=\partial_{\mu}\partial_{\nu}h+\cdots\,.&\nonumber\eea{no-number}
One can substitute the right hand sides of the constraints~(\ref{trace1})--(\ref{div-cons}) in Eq.~(\ref{del-G10}) to write down the true dynamical
equations. It turns out that the equations of motion for the longitudinal modes completely decouple. They have the following form~\changed{\cite{footnote5}}:
\beq\begin{aligned} \left(\partial^2-m^2\right)h_{vi}&=Y_{ij}\partial_v^2h_{vj}+Y_i\partial_vh_{vv},\\
\left(\partial^2-m^2\right)h_{vv}&=Z_i\partial_v^3h_{vi}+Z\partial_v^2h_{vv}\,,\end{aligned}\eeq{eom12}
where we have defined the following operators:
\beq\begin{aligned} Y_{ij}=2(\hat\alpha-1)F\delta_{ij}-\tfrac{1}{\hat{m}^2}\left(F_{,\:\!ij}
+F_{,\:\!j}\partial_i\right),~~~~~~&\\
\!\!\!Y_i=2\hat\alpha\left(F_{,\:\!i}+F\partial_i\right)+2F_{,\:\!i} +\tfrac{1}{\hat{m}^2}
\left(F_{,\:\!ij}+F_{,\:\!j}\partial_i\right)\partial_j,&\\
Z_i=-\tfrac{1}{\hat{m}^2}F_{,\:\!i}\,,\quad
Z=4\left(\hat\alpha-\tfrac{1}{2}\right)F+\tfrac{1}{\hat{m}^2}F_{,\:\!i}\partial_i\,.~~&\end{aligned}\eeq{part101}
The transverse derivatives of $F$ are given, in terms of the unit transverse-position vector $\vec n\equiv\vec{x}/|\vec{x}|$,
as
\beq\begin{aligned} F_{,\:\!i}=-mFn_i\,\tfrac{K_1(m|\vec{x}|)}{K_0(m|\vec{x}|)},~~~~~~~~~~~~~~~&\\
F_{,\:\!ij}=m^2F\!\left[n_in_j+\tfrac{K_1(m|\vec{x}|)}{m|\vec{x}|\,K_0(m|\vec{x}|)}\!\left(2n_in_j-\delta_{ij}\right)\right]\,.&\end{aligned}\eeq{identity}

%%%%%%%%%%%%%%%%%%%%%%%%%%%%%%%%%
\section{SHAPIRO TIME DELAY / ADVANCE}
%%%%%%%%%%%%%%%%%%%%%%%%%%%%%%%%%

One of the classic tests of General Relativity is the Shapiro time delay~\cite{Shapiro:1964uw} suffered by a light ray while passing by a massive body.
We would like to compute this delay (or advance) for the longitudinal modes of the massive-gravity fluctuation upon crossing the sandwich wave. To this end,
we note that the general solutions of Eq.~(\ref{del-G10}) and the constraints ~(\ref{trace1})--(\ref{div-cons}) can be written as superpositions of
eigensolutions of the form:
\beq h_{\mu\nu}(u,v,\vec{x})=\tilde{h}_{\mu\nu}(u)\,e^{i\left(pv+\vec{q}\cdot\vec{x}\right)}\,,\eeq{Fourier}
where $p$ and $\vec q$ are the momenta in the $u$-direction and the transverse directions respectively.

Note that $\vec q=\vec q(u)$ since the probe will experience a radial impulse in the transverse plane during the course of the sandwich wave, $u\in[-\lambda,\lambda]$.
Let $\vec q_{_-}$ and $\vec q_{_+}$ be the incoming and outgoing transverse momenta respectively.
We denote by $\vec b$ the impact parameter vector (in the transverse plane) at $u=-\lambda$.
The unit vector along this direction is $\vec e\equiv\vec{b}/b$, where $b=|\vec{b}|$. We choose $\vec q_{_-}$ to be aligned with $\vec{b}$, i.e.,
$\vec q_{_-}=q_{_-}\vec e$~\,with~\,$q_{_-}\!>0$.

We will consider the following regime of parameters:
\beq \Lambda\gtrsim\frac{1}{\lambda}\gg p\gg q_{_-}\gg\frac{1}{b}\gg m\,.\eeq{parameter-regime}
The above parametric relations may very well be accommodated because the separation between the scales $\Lambda$ and $m$ is huge $\!\sim\!\sqrt[3]{M_\textrm{P}/m}$.
The condition $q_{_-}b\gg1$ ensures that the probe is far away from $\vec x=0$.
For simplicity of analysis, we take the particle to be ultrarelativistic with $p\gg q_{_-}\gg m$, but all momenta are much smaller than $\Lambda$.
On the other hand, the sandwich wave is chosen to be very thin compared to the length scales characterizing the probe: $\lambda p\ll1$,
but thick enough to be ``seen'' in the effective theory: $\lambda\Lambda\gtrsim1$. Finally, the choice of a small impact parameter, $mb\ll1$,
amplifies the effects of the sandwich wave on the probe.

While the probe particle is passing through the sandwich wave, its transverse position $\vec x$ will change slightly: $|\vec x-\vec b|\lesssim\lambda$,
$|\vec{n}-\vec{e}\,|\lesssim\lambda/b$. We will neglect these small changes. The radial impulse deflects the particle but keeps $\vec q(u)$ aligned
with $\vec e$: $\vec q(u)=q(u)\vec e$. Note that $q(u)$ remains positive and small compared to $p$. To see this, let us use the deflection formula~(A.36)
of Ref.~\cite{Dray:1984ha}, which is a valid approximation because the sandwich wave is thin. With $E\sim M_\textrm{P}^2\lambda$ and $\vec q_{_+}\equiv q_{_+}\vec e$,
we can write
$(q_{_-}/p)-(q_{_+}/p)\sim\lambda/b$.
Given the separation of scales~(\ref{parameter-regime}), we conclude that $q_{_+}\!>0$~\,and~\,$q_{_+}\approx q_{_-}$. The same conclusion holds for $q(u)$
as it varies continuously.

Let us collectively denote the longitudinal modes as $\{\Phi_I(u)\}$ with $I=1,2,3$, defined as
\beq \Phi_1=e_i\tilde{h}_{vi}\,,\qquad\Phi_2=\varepsilon_{ij}e_i\tilde{h}_{vj}\,,\qquad\Phi_3=\tilde{h}_{vv}\,,\eeq{modes-defined}
where $\varepsilon_{ij}$ is the Levi--Civita symbol in the transverse plane. Now, plugging the expressions~(\ref{Fourier}) into Eqs.~(\ref{eom12})
and using the redefinitions~(\ref{modes-defined}) results in the following first-order coupled differential equations:
\beq \left(\partial_u-ip\gamma\right)\Phi_I(u)=ipA(u)\mathcal{M}_{IJ}\Phi_J(u),\eeq{DE-0}
where $\gamma\equiv\tfrac{1}{4}(q^2+m^2)/p^2$, and the $3\hspace{-0pt}\times\hspace{-0pt}3$ matrix $\mathcal{M}$ contains the functions $K_0(mb)\equiv k_0$ and $K_1(mb)\equiv k_1$
in the following non-zero components:
\beq\begin{aligned} \!\!\!\!\mathcal{M}_{11}&=\tfrac{\hat\alpha-3}{6}\,k_0+\tfrac{2\hat\alpha(1-iqb)}{3mb}\,k_1\,,\\
\mathcal{M}_{13}&=-\tfrac{7\hat\alpha q}{6p}\,k_0-\tfrac{4\hat\alpha q+3i(\hat\alpha+1)m^2b-4i\hat\alpha q^2b}{6pmb}\,k_1\,,\\
\mathcal{M}_{22}&=-\tfrac{\hat\alpha+1}{2}\,k_0-\tfrac{2\hat\alpha}{3mb}\,k_1\,,\\
\mathcal{M}_{31}&=-\tfrac{2i\hat\alpha p}{3m}\,k_1\,,\quad
\mathcal{M}_{33}=-\tfrac{2\hat\alpha+1}{2}\,k_0+\tfrac{2i\hat\alpha q}{3m}\,k_1\,.\end{aligned}\eeq{m33}
Let the eigenvalues of $\mathcal M$ be $\mu_I$. Explicit computation shows that they are independent of both $p$ and $q$. The matrix $\mathcal P$ composed of the eigenvectors
of $\mathcal M$ is $u$-dependent, but this dependency
is as small as $q(u)/p$. Then, in terms of the modes $\Phi'_I\equiv\mathcal{P}^{-1}_{IJ}\Phi_J$, Eqs.~(\ref{DE-0}) are approximately diagonal, and hence
can be integrated to
\beq \Phi'_I(+\lambda)\approx\Phi'_I(-\lambda)e^{ip\int_{-\lambda}^{+\lambda}\mathrm{d}u\,\left(\gamma+\mu_IA(u)\right)}\,.\eeq{integrated}
\changed{Note that the dynamical equations for the diagonalized modes $\Phi'_I$ are second order in $p$, despite the fact that in Eqs.~\eqref{eom12} there appear $\partial_v^3$-terms
(they arise from the mixing of modes).}

The integral in Eq.~\eqref{integrated} is to be understood as the shift in the $v$-coordinate suffered by the $I$-th mode upon crossing the sandwich
wave~\cite{Camanho:2014apa}. To find the shift relative to massless propagation in flat space, we write the relevant terms originating from $\gamma$:
\beq \Delta\gamma=\tfrac{1}{4}m^2/p^2+\tfrac{1}{4}(q^2-q_{_-}^2)/p^2\,.\eeq{gamma-break}
The first piece comes from the non-zero graviton mass, whereas the second from the non-zero curvature.
\begin{center}
\begin{figure}[h]
\hspace*{-25pt}
\begin{minipage}{.2\textwidth}
	\includegraphics{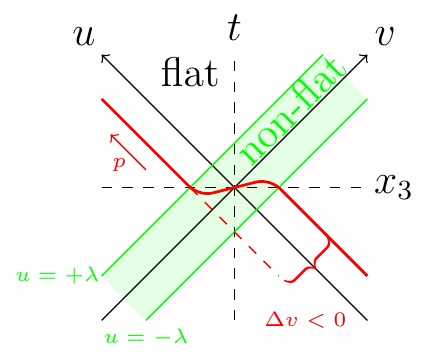}
\end{minipage}
\begin{minipage}{.05\textwidth}
\hphantom{hhhh}
\end{minipage}
\begin{minipage}{.2\textwidth}
	\includegraphics{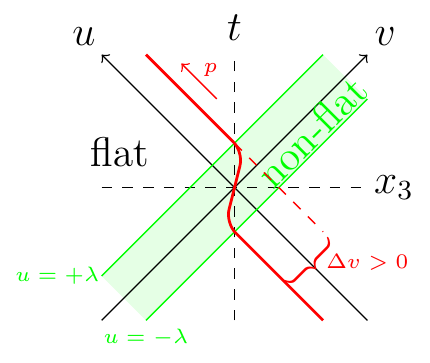}
\end{minipage}
\caption{Upon crossing the sandwich wave the probe undergoes a time advance (left) or a time delay (right).}
\end{figure}
\end{center}
%\vspace*{-20pt}
The eigenvalues $\mu_I$ are independent of $p$ and $q$. For small impact parameter $mb\ll1$ they reduce to
\beq \mu_1=\tfrac{2\hat\alpha}{3m^2b^2}\,,~~\mu_2=-\tfrac{2\hat\alpha}{3m^2b^2}\,,~~\mu_3=\tfrac{2\hat\alpha+1}{2}\ln(mb)\,,\eeq{eigenvalues}
and dominate over the $\Delta\gamma$-contributions \cite{footnote1.5}. Then, the $v$-shift relative to flat-space massless propagation reads
\beq \Delta v_I\equiv \int_{-\lambda}^{+\lambda}\mathrm{d}u\left(\Delta\gamma+A(u)\mu_I\right)\approx \lambda\mu_I\,.\eeq{v-shift}
A positive shift corresponds to a time delay, whereas a negative $\Delta v$ corresponds to a time advance (see Fig.~2).

Any value of $\hat\alpha$ yields a time advance for at least one of the modes. Since $\mu_1$ and $\mu_2$ have opposite signs, Eq.~(\ref{v-shift}) says that any non-zero value of $\hat\alpha$  will lead to a time advance either for $\Phi_1'$ or $\Phi_2'$. In that case, for $|\hat\alpha|\gtrsim m^2b^2$ the time advance is larger than the resolution time of
the effective theory: $|\Delta v|\gtrsim1/\Lambda$. When $\hat\alpha = 0$ both $\mu_1$ and $\mu_2$ are zero, but the third eigenvalue $\mu_3$ is negative because $1 \gg m b$~\cite{footnote2}. Then, the third mode $\Phi_3'$ undergoes a time advance $|\Delta v|>\mathcal{O}(\lambda)$, which is measurable in the effective theory.

%%%%%%%%%%%%%%%%%%%%%%%%%%%%%%%
\section{CLOSED TIME-LIKE CURVES}
%%%%%%%%%%%%%%%%%%%%%%%%%%%%%%%

The argument that time advances lead to close time-like curves (CTC) is standard. For the sake of self-containedness we just present the arrangements
of Appendix~G of Ref.~\cite{Camanho:2014apa}. Strictly speaking, one would need a refined version of the simplistic setup appearing in Fig.~3.
\tikzset{->-/.style={decoration={
  markings,
  mark=at position #1 with {\arrow{>}}},postaction={decorate}}}
\begin{center}
\begin{figure}[h]
\hspace*{-5pt}
\begin{minipage}{.2\textwidth}
	\includegraphics{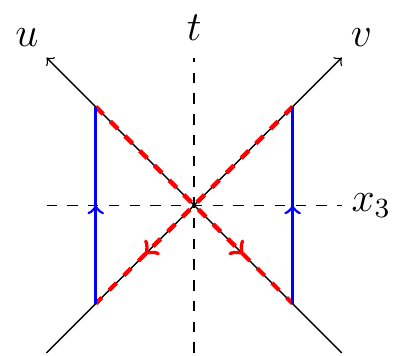}
\end{minipage}
\begin{minipage}{.05\textwidth}
\hphantom{hhhh}
\end{minipage}
\begin{minipage}{.2\textwidth}
	\includegraphics{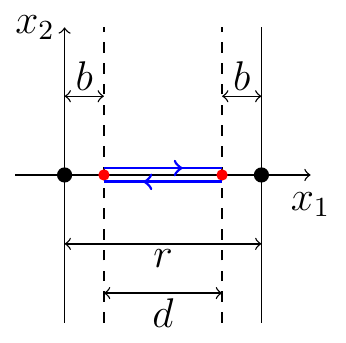}
\end{minipage}
\caption{Motion of a probe that follows a CTC, projected on the $u$-$v$ plane (left) and on the transverse plane (right).}
\end{figure}
\end{center}
We imagine two sandwich waves moving in opposite directions, centered respectively at $u=0$ and $v=0$, separated in the transverse plane
by a distance $r$. The probe crosses the waves one after the other, and \changed{for $\hat\alpha \neq 0$ either $\Phi'_1$ or $\Phi'_2$} acquires time advances $|\Delta v|=|\Delta u|\sim|\hat\alpha|\lambda/(m^2b^2)$. Note that there is a contribution to the time delay from the non-zero mass, since the probe travels a finite distance~$\sim|\Delta v|$. This contribution goes as $|\Delta v| (m^2 / p^2)$ and is therefore negligible in front of $|\Delta v|$. Right after each wave passes by, a mirror is needed to control the  motion in the transverse plane. The mirrors must be set in appropriate angles to counter deflections.

In between the two waves the probe travels a transverse distance $d=r-2b$. In order to form a CTC we would need $d\sim|\hat\alpha|\lambda/(m^2b^2)$.
We also require $d\gg1/m$, so that the waves have negligible overlap at $u=v=0$. These requirements combine into
$m^2b^2\ll|\hat\alpha|m\lambda$. In other words, the small numbers $\epsilon_1\equiv mb$ and
$\epsilon_2\equiv\lambda/b$ should be chosen such that $\epsilon_1/\epsilon_2\ll|\hat\alpha|$. With the present LIGO
bound~\cite{Abbott:2016blz} on the graviton mass one can maintain the separation of scales~(\ref{parameter-regime}) while making the ratio $\epsilon_1/\epsilon_2$ as small as $10^{-6}$.
Therefore, our argument leaves room for the following parameter region:
\beq |\hat\alpha|\sim|\alpha_3+\tfrac{1}{2}|\lesssim10^{-6}\,.\eeq{band}
This is in spirit the line $\alpha_3=-1/2$, reported in Eq.~(\ref{mainresult}). With improved bounds on the graviton mass, this region will only get narrower.

One still needs to see if the third mode $\Phi'_3$ can form CTCs in the parameter region~\eqref{band}. A similar analysis shows that this can happen if $\epsilon_1 |\ln\epsilon_1| \gg 1/\epsilon_2$.
It is easy to see that in the regime~(\ref{parameter-regime}) under consideration, the above parametric condition cannot be satisfied, so that CTCs may not be formed in the region~\eqref{band}.

%%%%%%%%%%%%%%%%%%%%%%%%%%%%%
\section{SUMMARY AND REMARKS}
%%%%%%%%%%%%%%%%%%%%%%%%%%%%%

Within the dRGT theory's validity regime we find that
\vspace{-2pt}
\begin{enumerate}
 	\item For all values of $\alpha_3$ and $\alpha_4$, physical modes of the theory undergo measurable time advances,
 	\item Outside the region~(\ref{mainresult}), CTCs can be formed.
\end{enumerate}

The time advances experienced by the longitudinal modes, relative to flat-space massless propagation, may themselves be regarded as a serious pathology~\cite{footnote7}. This is presumably an IR manifestation of the restrictions arising from requiring a sensible UV-completion~\cite{Adams:2006sv}. Indeed, in order for a theory to make sense in the UV, a sensible requirement would be that the asymptotic Lorentz invariance is respected, so that one can define an S-matrix.

Should the time advances alone not be considered a pathology of the theory, CTC formation would leave one only with the region of validity~(\ref{mainresult})~\cite{lastfoot}. 
In this regard it is interesting to look at the Cheung--Remmen parameter island (see Fig.~4), singled out by positivity constraints on
scattering amplitudes~\cite{Cheung:2016yqr}.
\begin{figure}[ht]
	\begin{center}
	\includegraphics{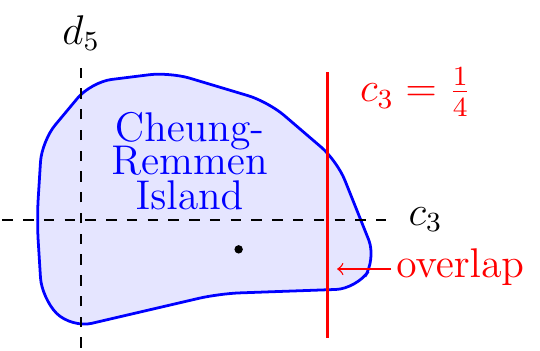}
	\end{center}
	%\vspace{-5pt}
	\caption{A cartoon of the Cheung-Remmen parameter island. The red line $c_3=1/4$ corresponds to our result $\alpha_3=-1/2$.}
\end{figure}
%\vspace{-5pt}
In the parameter plane of $\left(c_3,d_5\right)\equiv\left(-\alpha_3/2,-\alpha_4/4\right)$, our result corresponds to the line $c_3=1/4$,
and it rules out the minimal model represented by the black dot of Fig.~4 at $c_3=1/6$ and $d_5=-1/48$. The positivity constraints are necessary but not sufficient for the existence of a UV-completion. Our analysis, on the other hand, considers only one class of backgrounds allowed by the dRGT theory.

\vspace{12pt}
%%%%%%%%%%%%%%%%%%%%%%%%%%%%
\section*{ACKNOWLEDGEMENTS}
%%%%%%%%%%%%%%%%%%%%%%%%%%%%

We are thankful to P.~Creminelli, K.~Hinterbichler, Y.~Korovin, I.~R\'acz, S.~Theisen, A.~Waldron and A.~Zhiboedov for useful discussions and comments. This work is partially supported by the Deutsche Forschungsgemeinschaft (DFG) GZ: OT 527/2-1.  The research of GLG is supported by the Alexander
von Humboldt Foundation.

%%%%%%%%%%%%%%%%%%%%%%%%%%%


\begin{thebibliography}{100}
%%%%%%%%%%%%%%%%%%%%%%%%%%%

\bibitem{Fierz:1939ix}
  M.~Fierz and W.~Pauli,
  %``On relativistic wave equations for particles of arbitrary spin in an electromagnetic field,''
  Proc.\ Roy.\ Soc.\ Lond.\ A {\bf 173}, 211 (1939).
  %doi:10.1098/rspa.1939.0140
  %%CITATION = doi:10.1098/rspa.1939.0140;%%
  %1023 citations counted in INSPIRE as of 30 Sep 2016

\bibitem{deRham:2010ik}
  C.~de Rham and G.~Gabadadze,
  %``Generalization of the Fierz-Pauli Action,''
  Phys.\ Rev.\ D {\bf 82}, 044020 (2010)
  %doi:10.1103/PhysRevD.82.044020
  [arXiv:1007.0443 [hep-th]].
  %%CITATION = doi:10.1103/PhysRevD.82.044020;%%
  %595 citations counted in INSPIRE as of 01 Sep 2016

\bibitem{deRham:2010kj}
  C.~de Rham, G.~Gabadadze and A.~J.~Tolley,
  %``Resummation of Massive Gravity,''
  Phys.\ Rev.\ Lett.\  {\bf 106}, 231101 (2011)
  %doi:10.1103/PhysRevLett.106.231101
  [arXiv:1011.1232 [hep-th]].
  %%CITATION = doi:10.1103/PhysRevLett.106.231101;%%
  %759 citations counted in INSPIRE as of 01 Sep 2016

\bibitem{Boulware:1973my}
  D.~G.~Boulware and S.~Deser,
  %``Can gravitation have a finite range?,''
  Phys.\ Rev.\ D {\bf 6}, 3368 (1972).
  %doi:10.1103/PhysRevD.6.3368
  %%CITATION = doi:10.1103/PhysRevD.6.3368;%%
  %679 citations counted in INSPIRE as of 30 Sep 2016

\bibitem{Hassan:2011hr}
  S.~F.~Hassan and R.~A.~Rosen,
  %``Resolving the Ghost Problem in non-Linear Massive Gravity,''
  Phys.\ Rev.\ Lett.\  {\bf 108}, 041101 (2012)
  %doi:10.1103/PhysRevLett.108.041101
  [arXiv:1106.3344 [hep-th]];
  %%CITATION = doi:10.1103/PhysRevLett.108.041101;%%
  %435 citations counted in INSPIRE as of 01 Sep 2016
%\bibitem{Hassan:2011ea}
%  S.~F.~Hassan and R.~A.~Rosen,
  %``Confirmation of the Secondary Constraint and Absence of Ghost in Massive Gravity and Bimetric Gravity,''
  JHEP {\bf 1204}, 123 (2012)
  %doi:10.1007/JHEP04(2012)123
  [arXiv:1111.2070 [hep-th]].
  %%CITATION = doi:10.1007/JHEP04(2012)123;%%
  %275 citations counted in INSPIRE as of 30 Sep 2016

\bibitem{Velo:1972rt}
  G.~Velo,
  %``Anomalous behaviour of a massive spin two charged particle in an external electromagnetic field,''
  Nucl.\ Phys.\ B {\bf 43}, 389 (1972).
  %doi:10.1016/0550-3213(72)90027-2
  %%CITATION = doi:10.1016/0550-3213(72)90027-2;%%
  %47 citations counted in INSPIRE as of 30 Sep 2016

\bibitem{Henneaux:2013vca}
  M.~Henneaux and R.~Rahman,
  %``Note on Gauge Invariance and Causal Propagation,''
  Phys.\ Rev.\ D {\bf 88}, 064013 (2013)
  %doi:10.1103/PhysRevD.88.064013
  [arXiv:1306.5750 [hep-th]].
  %%CITATION = doi:10.1103/PhysRevD.88.064013;%%
  %6 citations counted in INSPIRE as of 30 Sep 2016

\bibitem{Camanho:2014apa}
  X.~O.~Camanho, J.~D.~Edelstein, J.~Maldacena and A. Zhiboedov,
  %``Causality Constraints on Corrections to the Graviton Three-Point Coupling,''
  JHEP {\bf 1602}, 020 (2016)
  %doi:10.1007/JHEP02(2016)020
  [arXiv:1407.5597 [hep-th]].
  %%CITATION = doi:10.1007/JHEP02(2016)020;%%
  %107 citations counted in INSPIRE as of 30 Sep 2016

\bibitem{Deser:2012qx}
  S.~Deser and A.~Waldron,
  %``Acausality of Massive Gravity,''
  Phys.\ Rev.\ Lett.\  {\bf 110}, no. 11, 111101 (2013)
  %doi:10.1103/PhysRevLett.110.111101
  [arXiv:1212.5835 [hep-th]].
  %%CITATION = doi:10.1103/PhysRevLett.110.111101;%%
  %107 citations counted in INSPIRE as of 30 Sep 2016
  S.~Deser, K.~Izumi, Y.~C.~Ong and A.~Waldron,
  %``Problems of massive gravities,''
  Mod.\ Phys.\ Lett.\ A {\bf 30}, 1540006 (2015)
  %doi:10.1142/S0217732315400064
  [arXiv:1410.2289 [hep-th]].
  %%CITATION = doi:10.1142/S0217732315400064;%%
  %20 citations counted in INSPIRE as of 11 Oct 2016
  S.~Deser, M.~Sandora, A.~Waldron and G.~Zahariade,
  %``Covariant constraints for generic massive gravity and analysis of its characteristics,''
  Phys.\ Rev.\ D {\bf 90}, no. 10, 104043 (2014)
  %doi:10.1103/PhysRevD.90.104043
  [arXiv:1408.0561 [hep-th]].
  %%CITATION = doi:10.1103/PhysRevD.90.104043;%%
  %28 citations counted in INSPIRE as of 11 Oct 2016

\bibitem{Hassan}
  S.~F.~Hassan and M.~Kocic,
  %``On the local structure of spacetime in ghost-free bimetric theory and massive gravity,''
  arXiv:1706.07806 [hep-th].
  %%CITATION = ARXIV:1706.07806;%%
  %4 citations counted in INSPIRE as of 19 Sep 2017
  
\bibitem{Bondi:1958aj}
  H.~Bondi, F.~A.~E.~Pirani and I.~Robinson,
  %``Gravitational waves in general relativity. 3. Exact plane waves,''
  Proc.\ Roy.\ Soc.\ Lond.\ A {\bf 251}, 519 (1959).
  %doi:10.1098/rspa.1959.0124
  %%CITATION = doi:10.1098/rspa.1959.0124;%%
  %132 citations counted in INSPIRE as of 30 Sep 2016

\bibitem{footnote1}
The situation is analogous to having a macroscopic (super-Planckian) black hole
in General Relativity.

\bibitem{Bernard:2014bfa}
  L.~Bernard, C.~Deffayet and M.~von Strauss,
  %``Consistent massive graviton on arbitrary backgrounds,''
  Phys.\ Rev.\ D {\bf 91}, no. 10, 104013 (2015)
  %doi:10.1103/PhysRevD.91.104013
  [arXiv:1410.8302 [hep-th]].
  %%CITATION = doi:10.1103/PhysRevD.91.104013;%%
  %13 citations counted in INSPIRE as of 02 Sep 2016
\bibitem{footnote5}
Notice the terms of order three in derivatives on the right-hand side of~(\ref{eom12}). They appear precisely because the scalar constraint~(\ref{trace2}),
which contains derivatives, has been used to eliminate the combination $h_{11} + h_{22}$.

\bibitem{Shapiro:1964uw}
  I.~I.~Shapiro,
  %``Fourth Test of General Relativity,''
  Phys.\ Rev.\ Lett.\  {\bf 13}, 789 (1964).
  %doi:10.1103/PhysRevLett.13.789
  %%CITATION = doi:10.1103/PhysRevLett.13.789;%%
  %294 citations counted in INSPIRE as of 03 Oct 2016

\bibitem{Dray:1984ha}
  T.~Dray and G.~'t Hooft,
  %``The Gravitational Shock Wave of a Massless Particle,''
  Nucl.\ Phys.\ B {\bf 253}, 173 (1985).
  %doi:10.1016/0550-3213(85)90525-5
  %%CITATION = doi:10.1016/0550-3213(85)90525-5;%%
  %217 citations counted in INSPIRE as of 01 Sep 2016

\bibitem{footnote1.5}
Recall that the expressions \eqref{eigenvalues} hold only for small $m b$. Looking at the exact expressions for the eigenvalues, one can check that all three of them go to zero for large values of $m b$, as they should.
\bibitem{footnote2}
When $\hat{\alpha}\neq 0$, the right-hand side of the scalar constraint~\eqref{trace2} is non-zero, which will always go along with causality violation (see also~\cite{Deser:2012qx}). Here we find that even if the trace of the graviton field is zero on-shell, the system exhibits time advances.

\bibitem{Abbott:2016blz}
  B.~P.~Abbott {\it et al.} [LIGO Scientific and Virgo Collaborations],
  %``Observation of Gravitational Waves from a Binary Black Hole Merger,''
  Phys.\ Rev.\ Lett.\  {\bf 116}, no. 6, 061102 (2016)
  %doi:10.1103/PhysRevLett.116.061102
  [arXiv:1602.03837 [gr-qc]].
  %%CITATION = doi:10.1103/PhysRevLett.116.061102;%%
  %666 citations counted in INSPIRE as of 13 Sep 2016

\bibitem{footnote7}
The absence of time advances has been used before in the literature, as a legitimate and necessary requirement for a theory to be healthy \cite{DAppollonio:2015fly}.
These are global time advances that should be understood as a violation of the asymptotic causal structure, i.e., Lorentz symmetry. Strictly speaking, the pp-wave
background does break asymptotic Lorentz invariance. Such a background, however, ought to be considered as an idealization of a coherent bunch of finite energy wave packets.
In a more realistic setting, the wave packets would have to get focused from infinity to form a mildly deformed pp-wave, and then disperse after
some time. Thus, instead of an energy source that remains concentrated in a small region from null past to null future, we would have ``radiation'' coming from null
past to form a transitory pp-wave before dispersing. In this way the energy would be dispersed both in the past and the future, and the solution would respect
asymptotic Lorentz invariance. For the sake of simplicity one may just analyze the transient pp-wave, and this is enough for our pourpose. Given this,
the light cone at infinity is fixed, and so the time advances we found correspond to causally connecting spacelike-separated points. Since the pp-wave vanishes
before and after $u=\pm\lambda$, a mode propagating from asymptotic infinity to asymptotic infinity would undergo the same time advances as found above.
These time advances are problematic if one uses the notion of causality introduced by Gao and Wald~\cite{Gao:2000ga}: one cannot send signals faster than what is allowed by 
the asymptotic causal structure of the spacetime (see also~\cite{Olum:1998mu}). This notion has already been used in Ref.~\cite{Camanho:2014apa}.

\bibitem{Adams:2006sv}
  A.~Adams, N.~Arkani-Hamed, S.~Dubovsky, A.~Nicolis and R.~Rattazzi,
  %``Causality, analyticity and an IR obstruction to UV completion,''
  JHEP {\bf 0610}, 014 (2006)
  %doi:10.1088/1126-6708/2006/10/014
  [hep-th/0602178].
  %%CITATION = doi:10.1088/1126-6708/2006/10/014;%%
  %370 citations counted in INSPIRE as of 03 Oct 2016

\bibitem{lastfoot}
In our setup, forming CTCs requires ``mirrors'' that reflect the probe. Therein, it is tacitly assumed that such ``reflections'' do not introduce significant time delays
that would wash out the time advances. The dRGT theory could, in principle, dynamically protect itself against formation of CTCs (see for example~\cite{Hassan}).

\bibitem{Cheung:2016yqr}
  C.~Cheung and G.~N.~Remmen,
  %``Positive Signs in Massive Gravity,''
  JHEP {\bf 1604}, 002 (2016)
  %doi:10.1007/JHEP04(2016)002
  [arXiv:1601.04068 [hep-th]].
  %%CITATION = doi:10.1007/JHEP04(2016)002;%%
  %6 citations counted in INSPIRE as of 03 Oct 2016

\bibitem{DAppollonio:2015fly}
  G.~D'Appollonio, P.~Di Vecchia, R.~Russo and G.~Veneziano,
  %``Regge behavior saves String Theory from causality violations,''
  JHEP {\bf 1505} (2015) 144
  doi:10.1007/JHEP05(2015)144
  [arXiv:1502.01254 [hep-th]].
  %%CITATION = doi:10.1007/JHEP05(2015)144;%%
  %19 citations counted in INSPIRE as of 13 Jan 2017

\bibitem{Gao:2000ga}
  S.~Gao and R.~M.~Wald,
  %``Theorems on gravitational time delay and related issues,''
  Class.\ Quant.\ Grav.\  {\bf 17}, 4999 (2000)
  %doi:10.1088/0264-9381/17/24/305
  [gr-qc/0007021].
  %%CITATION = doi:10.1088/0264-9381/17/24/305;%%
  %92 citations counted in INSPIRE as of 14 Feb 2017

\bibitem{Olum:1998mu}
  K.~D.~Olum,
  %``Superluminal travel requires negative energies,''
  Phys.\ Rev.\ Lett.\  {\bf 81}, 3567 (1998)
  %doi:10.1103/PhysRevLett.81.3567
  [gr-qc/9805003].
  %%CITATION = doi:10.1103/PhysRevLett.81.3567;%%
  %65 citations counted in INSPIRE as of 29 May 2017


\end{thebibliography}
\end{document}